\documentclass[12pt,reqno]{amsart}
\usepackage[headings]{fullpage}
\usepackage{amssymb,amsmath,mathtools,bbm,tikz,tikz-cd,color,mathrsfs, bm,appendix}
\usepackage[bookmarks=true,%
colorlinks=true,%
linkcolor=blue,%
citecolor=blue,%
filecolor=blue,%
menucolor=blue,%
urlcolor=blue,%
breaklinks=true]{hyperref}

\newtheorem{theorem}{Theorem}[section]
\theoremstyle{definition}
\newtheorem{proposition}[theorem]{Proposition}
\newtheorem{lemma}[theorem]{Lemma}
\newtheorem{definition}[theorem]{Definition}
\newtheorem{remark}[theorem]{Remark}
\newtheorem{corollary}[theorem]{Corollary}

\def\be{\begin{equation}}
	\def\ee{\end{equation}}

\def\a{\alpha}
\def\b{\beta}
\def\g{\gamma}

\def\e{\epsilon}

\def\th{\theta}
\def\si{\sigma}

\def\BZ{\mathbbm Z}
\def\BQ{\mathbbm Q}
\def\BR{\mathbbm R}
\def\BC{\mathbbm C}

\def\calC{\mathcal C}
\def\calH{\mathcal H}
\def\calI{\mathcal I}

\def\sX{\mathsf{X}}

\def\id{\mathrm{id}}
\def\End{\mathrm{End}}
\def\Hom{\mathrm{Hom}}

\newcommand{\lcoev}{\stackrel{\longleftarrow}{\operatorname{coev}}}
\newcommand{\lev}{\stackrel{\longleftarrow}{\operatorname{ev}}}
\newcommand{\rev}{\stackrel{\longrightarrow}{\operatorname{ev}}}
\newcommand{\rcoev}{\stackrel{\longrightarrow}{\operatorname{coev}}}
\def\mt{\mathsf{tr}}

\def\md{\mathsf{d}}
\def\br#1{ \{ #1 \}}
\newcommand\qbin[2]{\begin{bmatrix}#1\\#2
	\end{bmatrix}_q}

\newcommand{\ket}[1]{|#1\rangle}

\def\bW{\bm{W}}
\def\bV{\bm{V}}

\def\p#1{\ensuremath{|{#1}|}}

\def\Uq{{U}_q}
\def\UqGH{{U}_q^H(\mathfrak{sl}(2|1))}

\def\Hom{\mathrm{Hom}}

\def\cat{\calC}
\def\catsl{\calC^{\mathfrak{sl}(2|1)}}
\def\catslR{\calC^{\mathfrak{sl}(2|1)}_{\mathbb{R}}}
\def\idx{\calI}
\def\hilb{\calH}
\def\sl21{\mathfrak{sl}(2|1)}
\def\trr{\mathrm{tr}_R}
\def\tc{\widetilde{c}}

\begin{document}
	
	\title[Universal braiding in Links--Gould topological quantum computation]{An example of universal braiding in Links--Gould topological quantum computation}
	
	\author[M. Harper]{Matthew Harper}
	\address{Department of Mathematics, Michigan State University, East Lansing, Michigan, USA}
	\email{mrhmath@proton.me}
	
	\author[B.-M. Kohli]{Ben-Michael Kohli}
	\address{Section de Math\'ematiques, Universit\'e de Gen\`eve \\
		rue du Conseil-G\'en\'eral 7-9, 1205 Gen\`eve, Switzerland}
	\email{bm.kohli@protonmail.ch}

	
	\date{November 2, 2026}
	
	\begin{abstract}
		We construct a universal quantum computational encoding via braiding in Links--Gould fusion spaces associated to $U_q^H(\mathfrak{sl}(2|1))$ where $q$ is on the unit circle and is not a root of unity. The construction uses only neglectons and is realized in the morphism spaces $\mathrm{Hom}_{U_q^H}(V(0,\alpha),V(0,-\frac{1}{2})^{\otimes 2N}\otimes V(0,\alpha)).$ Under explicit positivity and arithmetic constraints these spaces admit a $2^N$-dimensional computational subspace with a positive definite Hermitian form. We prove single-qubit density in $\mathrm{U}(2,1)$ and approximate arbitrary unitary operators on $N$-qubit computational subspaces with arbitrarily small leakage. The calculations use explicit inclusions, projections, and Hermitian pairings without requiring $F$-matrices.  
	\end{abstract}
	\maketitle
	
	\section{Introduction}
	Inspired by recent work of \cite{IKLS25,IKLS26}, which established universal quantum computation via braiding in a non-semisimple extension of the Ising anyon model, we investigate an analogous construction for quantum $\mathfrak{sl}(2|1)$. The usual Ising category is the semisimplification of the category of quantum $\mathfrak{sl}_2$ representations at $q=e^{i\pi/4}$. The non-semisimple extension uses representations of the unrolled restricted quantum group, which includes finite-dimensional modules with continuous highest-weight parameters and vanishing quantum dimensions. These objects are referred to as neglectons in \cite{IKLS25,IKLS26}, and define the ADO link invariants in the modified Reshetikhin--Turaev construction \cite{ADO, GPT, GKPM11}. 

    The Hermitian structures developed in \cite{GLPS22,GLPS24} equip the associated fusion path spaces with nondegenerate pairings compatible with the braiding. By considering a positive definite subspace of $\Hom(\bm{\a},\bm{\a}\otimes \bm{\si}^{2N})$, the authors of \cite{IKLS25,IKLS26} exhibit computational universality of the braid group action. In this notation, $\bm{\a}$ is a typical $4$-dimensional representation of highest weight $\a$ and $\bm{\si}$ is the 2-dimensional defining representation. The encoding restricts to a $2^N$-dimensional subspace which is positive definite for particular values of $\a$.
    
	In the present article, we prove universal braiding within the category of quantum $\mathfrak{sl}(2|1)$ modules, building on the representation theoretic machinery of \cite{GHKKW}. We consider the Kac modules $V(n,\a)$ with highest-weight parameters $(n,\a)\in \BZ_{\geq0}\times \BC$ which compute the colored Links--Gould polynomials \cite{Links-Gould, DWKL, DW01,GHKKW}.  We construct our encoding from the morphism spaces $\Hom_{U_q^H}(\bV,\bW^{\otimes 2N}\otimes \bV)$ where 
    \begin{align}
        \bV=V(0,\a)\qquad \mbox{and}\qquad \bW=V(0,-\frac{1}{2}).
    \end{align}

    Our qubit encoding is similar in spirit to \cite{IKLS25,IKLS26}, but has several distinguishing features. First, our category is associated to a rank-two quantum supergroup and the quantum parameter $q=e^{i\th}$ remains variable and is not a root of unity. The key objects $\bV$ and $\bW$ in the construction are both neglectons and the tensor product $\bW^{\otimes 2}$ is non-semisimple, while $\bW^{\otimes 2}\otimes \bV$ is semisimple for generic $\a$. Moreover, we do not need to compute $F$-symbols, also called $6j$-symbols, which would involve non-semisimple tensor products. Instead, we compute braid matrices by explicitly calculating the composition of the braiding with inclusions and projections in the right-associated tensor product.
    
	\subsection{Overview of the construction and main result}
	The anyons in our theory are based on the Kac modules $V(n,\a)$ of $U^H_q(\mathfrak{sl}(2|1))$ of dimension $4(n+1)$. We discuss their construction in Section \ref{sec.uq}. Fix $q=e^{i\th}$ and $\a\in \BR$. Throughout the article we assume the arithmetic constraint:
	\begin{align}\label{eq.generic}
		1,\qquad \frac{\th}{2\pi}, \qquad \frac{\a\th}{2\pi} \qquad\mbox{are $\BZ$-linearly independent.}
	\end{align}
	In particular, $q$ is not a root of unity and weight labels on fusion paths under fusion with $\bW$ are typical. Under typicality assumptions, Kac modules satisfy the fusion rule
	\begin{equation}\label{eq.fusion}
		\begin{aligned}
			V(0,\b)\otimes V(n,\a)\cong &~
			V(n,\a+\b)\oplus V(n,\a+\b+1)
			\\&\oplus V(n+1,\a+\b)\oplus (1-\delta_{n,0})V(n-1,\a+\b+1).
		\end{aligned}
	\end{equation}
	The notation $(1-\delta_{n,0})$ in the last term indicates that this summand is dropped if $n=0$. Let us label each of the four summands $s$, $u$, $r$, and $l$. Summands within iterated tensor products with $V(0,\b)$ are in bijection with paths in these labels.
	
	The key observation in defining our qubit space is that there exist cycles in path space which simultaneously preserve the genericity assumption necessary for \eqref{eq.fusion}. More precisely, observe
	\begin{equation}\label{eq.cycle}
		\begin{aligned}
			V(0,\a)\xmapsto[V(0,\b)\otimes -]{\mbox{$s$}} &V(0,\a+\b)
			\xmapsto[V(0,\b)\otimes -]{\mbox{$u$}} V(0,\a+2\b+1)
			\\
			V(0,\a)\xmapsto[V(0,\b)\otimes -]{\mbox{$u$}} &V(0,\a+\b+1)
			\xmapsto[V(0,\b)\otimes -]{\mbox{$s$}}V(0,\a+2\b+1)
			\\
			V(0,\a)\xmapsto[V(0,\b)\otimes -]{\mbox{$r$}} &V(1,\a+\b)
			\xmapsto[V(0,\b)\otimes -]{\mbox{$l$}}V(0,\a+2\b+1)
		\end{aligned}
	\end{equation}
	are all valid tensor product projections which return to $V(0,\a)$ when $\b=-1/2$. 
    
    Now define for each $N\geq 0$,
	\begin{equation}
		\begin{aligned}
			\hilb_N=\Hom(\bV,  \bW^{2N}\otimes \bV)
			\qquad \mbox{and}
			\qquad
			\hilb^{\mathrm{comp}}_N=\mathrm{span}\{i_w\mid w\in \{su,us\}^N\}\subset \hilb_N.
		\end{aligned}
	\end{equation}
	Our computation space $\hilb^{\mathrm{comp}}$ encodes $N$ qubits and has dimension $2^N$. It has a basis indexed by computational words (fusion paths) $w\in \{su,us\}^N$ but it is not closed under the braid group action. Terms which map out of the computational space are referred to as leakage.

    Leading to our result, in Section \ref{sec.Herm}, we construct an explicit Hermitian pairing on all typical Kac modules with real weights. Let $\catsl_\mathrm{Herm}$ denote the category of real-weight $U_q^H(\mathfrak{sl}(2|1))$-modules generated by tensor products of simple Hermitian objects and their duals. We prove the following theorem using \cite{GLPS24}.

    \begin{theorem}\label{thm.herm}
        The category $\catsl_\mathrm{Herm}$ is a Hermitian ribbon category.
    \end{theorem}

    The Hermitian structure defines pairings on morphism spaces whose source is a simple module. This endows the total encoding space $\hilb_N$ with a nondegenerate Hermitian form. We will see that for the specified parameters, the Hermitian form on $\hilb_1$ is indefinite, while the restriction to $\hilb^{\mathrm{comp}}_1$ is positive definite. Positive definiteness holds for arbitrary numbers of qubits. We then prove braid universality in Section \ref{sec.tqc}.
    
	\begin{theorem}\label{thm.univ}
		Suppose $0<\th<\frac{\pi}{3}$ and $1<\a<\frac{\pi}{\th}-2$ satisfy \eqref{eq.generic}. For each $N\geq 1$, braiding via the $R$-matrix implements universal computation on $\hilb_N^\mathrm{comp}$, that is, every unitary operator can be approximated with arbitrarily small error and leakage.
	\end{theorem}
    
    Our proof for the $\mathfrak{sl}(2|1)$ theory follows the key steps of \cite{IKLS25}. We first prove single-qubit density in $\mathrm{U}(2,1)$ using a Lie-theoretic argument. A calculation using \texttt{Python/SymPy} was used to compute the braiding on the fusion-path basis, the code is available at \cite{LGTQCGithub}. We then show that any unitary operator on the $2$-qubit computational space can be approximated. The orbit of $\hilb_2^{\mathrm{comp}}$ under a particular subgroup closes to a larger invariant space $\overline{\hilb}_2^{\mathrm{comp}}\subseteq \hilb_N$ which decomposes as two parameter shifts of the single-qubit encoding and two one-dimensional representations. From this decomposition and single-qubit universality, We are able to construct an entangling $CZ$ gate on the computational space. Naturality of the braiding implies these results hold on any single qubit or consecutive pair of qubits. The universality theorem of \cite[Thm.1.3]{BrylinskiBrylinski} then proves our result.

    Given that the Links--Gould polynomials are known to not be a function of the colored Jones polynomials \cite{HK:JonesLG}, which are related to the ADO polynomials \cite{Willetts}, as well as the known specialization of Links--Gould to $\mathrm{ADO}_3$ \cite{GHKKST}, it would be interesting to compare the complexity and efficiency of our encoding with the encoding of \cite{IKLS25,IKLS26} and of other ADO/non-semisimple $\mathfrak{sl}_2$ anyon theories.  
    
	The fusion inclusions, Hermitian pairings, and braid actions considered here provide a foundation for future investigations on encodings in similar path spaces. Moreover, it would be interesting to see whether universality holds for $(\th,\a)$ which are outside the imposed parameter region or fail the arithmetic constraint following ideas of \cite{IKLS26}. 

    \subsection*{Acknowledgements}
    MH was partially supported through the NSF-RTG grant \#DMS-2135960.
	
	\section{Representation theory for \texorpdfstring{$\UqGH$}{UqHsl(2|1)}}
	\label{sec.uq}
	
	We assume the ground field is the complex numbers $\BC$ unless stated otherwise.
	Fix $\th\in \BR$ such that $\frac{\th}{2\pi}\notin\BQ$ and set $q=e^{i\th}$. In this way, $q$ lies on the unit circle and is not a root of unity. For $x\in\BC$, let 
	\begin{align}    
	\br{x}=q^x-q^{-x}\qquad \mbox{and} \qquad [x]=\frac{\br{x}}{\br{1}}.
	\end{align}
	\subsection{The unrolled quantum group}
	We briefly recall the notation and conventions for the unrolled quantum group of $\mathfrak{sl}(2|1)$ as used in \cite{GHKKW}.
	
	In the $\mathfrak{sl}(2|1)$ root system, we assume $\a_1$ is an even simple root and  $\a_2$ is an odd simple root. The associated Cartan data is $A=(A_{ij})=\begin{pmatrix} 2 &-1\\-1&0 \end{pmatrix}$. Write $\p x$ for the parity of $x$ and for homogeneous elements $x,y$: $[x,y]=xy-(-1)^{\p x \p y}yx$.
	
	\begin{definition}
		Let $\Uq=\UqGH$ denote the \emph{unrolled quantum group}. It is the unital superalgebra with even generators $H_i$, $K_i$, $K_i^{-1}$,
		$E_1$, $F_1$ for $i=1,2$ and odd generators $E_2$, $F_2$ and relations 
		\begin{equation}
			\label{eq:rels}
			\begin{aligned}
				[H_i,H_j]=[H_i,K_j^\pm]= [K_i,K_j]=0\,,\quad K_iK_i^{-1}=1\,,\qquad
				[E_i,F_j]=\delta_{ij}\frac{K_i-K_i^{-1}}{q-q^{-1}} \,,
				\\
				[H_i,E_j]={A_{ij}} E_j\,,
				\quad
				[H_i,F_j]=-{A_{ij}} F_j\,,
				\quad 
				K_iE_j=q^{A_{ij}}E_jK_i\,,\quad K_iF_j=q^{-A_{ij}}F_jK_i\,,
				\\
				E_2^2=F_2^2=0,\quad E_1^2E_2-[2]E_1E_2E_1+E_2E_1^2=0\,,
				\quad
				F_1^2F_2-[2]F_1F_2F_1+F_2F_1^2=0\,.
			\end{aligned} 
		\end{equation}
	\end{definition}
	
	There is a unique Hopf superalgebra structure on $\UqGH$ with counit $\epsilon$,
	coproduct $\Delta$ and antipode $S$ defined
	on generators by
	\be
	\label{eq:HopfStructure}
	\begin{aligned}
		\Delta(E_i)&=E_i\otimes K_i+1\otimes E_i\,,
		&S(E_i)&=-E_iK_i^{-1}\,,
		&\epsilon(E_i)&=0\,,
		\\
		\Delta(F_i)&=F_i\otimes 1+ K_i^{-1}\otimes F_i\,,
		&S(F_i)&=-K_iF_i\,,
		&\epsilon(F_i)&=0\,,
		\\
		\Delta(H_i)&=H_i\otimes 1+1\otimes H_i\,,
		&S(H_i)&=-H_i\,,
		&\epsilon(H_i)&=0\,,
		\\
		\Delta(K_i)&=K_i\otimes K_i\,,
		&S(K_i)&=K_i^{-1}\,,
		&\epsilon(K_i)&=1\,.
	\end{aligned}
	\ee
	
	We define the odd elements
	\be
	\label{E12F12}
	E_{12}=E_1E_2 - q E_2E_1, \qquad F_{12}=F_2F_1 - q^{-1} F_1F_2 \,.
	\ee
	
	\subsection{Categories of weight modules}
	\label{sub.catsl21}
	
	Let $\mathfrak{h}\subset\Uq$ denote the subalgebra of generated by $H_1$ and $H_2$. A $\Uq$-\emph{weight module} is a $\BZ/2\BZ$-graded finite-dimensional	$\UqGH$-module that is semisimple with respect to the action of $\mathfrak{h}$ and where $K_i=q^{H_i}$ for $i=1,2$. Let $\catsl$ denote the category whose objects are $\UqGH$-weight modules and morphisms being even (parity preserving) $\Uq$-linear maps. It is a $\BC$-linear abelian category with monoidal structure given by the tensor product of $\Uq$-weight modules. The braiding and ribbon structures are discussed in Subsection \ref{sub.braiding} below. 

    We will restrict our attention to the full subcategory $\catslR$ of $\catsl$  generated by objects with real weights $(n,\a)\in\BZ_{\geq 0}\times \BR$. This restriction will be necessary when discussing the Hermitian structures.
	
	For $(n,\a) \in \Lambda = \BZ_{\geq 0} \times \BC$ let $V(n,\a)$ denote the Kac module with highest weight $(n,\a)$~\cite{Kac:superalgebra}. We will assume the parity of the highest weight vector is $n\mod 2$ and will specify its parity $p$ otherwise by writing $V_p(n,\a)$. In this way, $V(n,\a)=V_n(n,\a)$. The dimension of $V(n,\a)$ is $4(n+1)$. Fix a highest weight vector $v_0\in V(n,\a)$, then a basis is given by the vectors
	\be \label{eq.basis}
	\begin{aligned}
		\{
		v_{i,j}=F_{i,j}v_0\mid i\in\{0,\dots, n\}, j\in\{0,1,2,3\}
		\}
	\end{aligned}
	\ee
	where for $i \in \BZ_{\geq0}$,
	\be
	\label{nonPBW}
	F_{i,0}=F_1^i \,, \qquad
	F_{i,1}=F_2F_1^i \,, \qquad
	F_{i,2}=F_1F_2F_1^i \,, \qquad
	F_{i,3}=F_2F_1F_2F_1^i.
	\ee
	Similar notation applies for the $E_{i,j}$.
	
	The action of generators in this basis is given as follows. 
	\be
	\label{eq.action}
	\begin{aligned}
		F_1v_{i,k}&=
		\begin{cases}
			v_{i+1,0} &k=0\\
			v_{i,2} &k=1\\
			[2]v_{i+1,2}-v_{i+2,1} &k=2\\
			v_{i+1,3} &k=3
		\end{cases} 
		\,,
		&
		E_1v_{i,k}&=
		\begin{cases}
			[i][n-i+1]v_{i-1,0} &k=0\\
			[i][n-i+1]v_{i-1,1} &k=1\\
			[n-2i+1]v_{i,1}+[i][n-i+1]v_{i-1,2} &k=2\\
			[i][n-i+1]v_{i-1,3} &k=3
		\end{cases} 
		\,,
		\\
		F_2v_{i,k}&=
		\begin{cases}
			v_{i,1} &k=0\\
			v_{i,3} &k=2\\
			0 &k=1,3
		\end{cases} 
		\,,
		&
		E_2v_{i,k}&=
		\begin{cases}
			0 &k=0\\
			[\a+i]v_{i,0} &k=1\\
			[\a+i]v_{i+1,0} &k=2\\
			[\a+i+1]v_{i,2}-[\a+i]v_{i+1,1} &k=3
		\end{cases} 
	\end{aligned}
	\ee
	where $v_{i,k}=0$ if $i<0$ or $i>n$. The Cartan generators act diagonally according to the formulas
	\be
	\begin{aligned}
		H_1v_{i,k}
		&=
		\begin{cases}
			(n-2i+k)v_{i,k} & k=0,1\\
			(n-2(i+1)+k-1)v_{i,k} & k=2,3
		\end{cases}\,,
		&&
		H_2v_{i,k}
		&=
		\begin{cases}
			(\a+i)v_{i,k} & k=0,1\\
			(\a+i+1)v_{i,k} & k=2,3
		\end{cases}
	\end{aligned}\,.
	\ee
	
	We say that the weight $(n,\a)$ is \emph{typical} if $[\a][n+1+\a] \neq 0$. The representation $V(n,\a)$ is irreducible if and only if its highest weight is typical \cite{GHKKW}. 
	
    The duality maps on $\catsl$ are given by
\be
\begin{aligned}
\lev_V(f \otimes v)&= f(v)\,, 
& \quad
\lcoev_V(1) &= \sum_i v_i \otimes v_i^{*}\,, 
\\
\rev_V(v \otimes f) &= 
(-1)^{\p f \p v}f(K_2^{-2} v)\,,
& \quad
\rcoev_V(1)&= \sum_i (-1)^{\p v_i}v_i^{*} \otimes K_2^{2} v_i
\end{aligned}
\ee
for any basis $\{v_i\}$ of $V\in\catsl$ and associated dual basis $\{v_i^*\}$ of
$V^*$. Taking the dual of a Kac module preserves typicality, $V(n,\a)^*\cong V(n,-n-\a-1)$.

    For any endomorphism $f\in\End_\cat(V\otimes V)$, set
\be
\begin{aligned}
  \trr(f) &= (\id_V \otimes \rev_V ) \circ (f \otimes \id_{V^*})
  \circ (\id_V \otimes \lcoev_V ) \in\End_\cat(V)\,.
\end{aligned}
\ee
	
	\subsection{Fusion rules}
	\label{sub.fusion}
	
	In this section we recall the tensor product decomposition of two typical modules
	using a formula of Geer-Patureau-Mirand~\cite[Lem.1.3]{Geer:multi}
	\be
	\label{V0n}
	\begin{aligned}
		V(0,\b)\otimes V(n,\a)    \cong~ &  V(n,\a+\beta) \oplus V(n,\a+\beta+1) \oplus
		V(n+1,\a+\beta) \\ & \oplus (1-\delta_{n,0}) V(n-1,\a+\beta+1) 
	\end{aligned}
	\ee
	Here, we assume that all highest weights are typical so that the following set of vectors are (nonzero) highest weight. We denote the vectors $v_{0,i}$ in $V(0,\b)$ by $w_i$.
	\be
	\label{highv}
	\begin{aligned}
		&(s)~V(n,\a+\b)  : \,\, \Delta(E_{0,3}F_{0,3})(w_0\otimes v_0), \\
		&(u)~V(n,\a+\b+1)  : \,\,  \Delta(E_{0,3}F_{0,3})(w_0\otimes F_2F_1F_2v_0), \\
		&(r)~V(n+1,\a+\b)  : \,\, \Delta(E_{0,3}F_{0,3})(w_0\otimes F_2v_0), \\
		&(l)~V(n-1,\a+\b+1)  : \,\,
		\Delta(E_{0,3}F_{0,3})(w_0\otimes ([n]F_1F_2-[n+1]F_2F_1)v_0)\,.
	\end{aligned}
	\ee
	One also obtains a set of highest weight vectors by following the method of \cite[Lem.3.4.8]{Anghel-Geer}. Our highest weight differ as we have different coproduct and nonsimple weight vector conventions. We exclude the $(l)$ term and set $v_{1,1}=0$ when $n=0$.
	\be
	\label{highvexplicit}
	\begin{aligned}
		&(s)~V(n,\a+\b)  : \,\, w_0\otimes v_0, \\
		&(u)~V(n,\a+\b+1)  : \,\, 
		w_0\otimes v_{0,3} +
        w_1\otimes \left(\frac{[\a]}{q^{\a+1}[\b]} v_{1,1}-\frac{[\a+1]}{q^{\a+1}[\b]}v_{0,2}\right)\\
        &\phantom{(u)~V(n,\a+\b+1)  : \,\, }
        +\frac{[\a+n+1]}{q^{\a+n+2}[\b]}w_2\otimes v_{0,1}
        +\frac{[\a][\a+n+1]}{q^{2\a+n+2}[\b][\b+1]} w_3\otimes v_0
		, \\
		&(r)~V(n+1,\a+\b)  : \,\, w_0\otimes v_{0,1}-\frac{[\a]}{q^\a[\b]} w_1\otimes v_0, \\
		&(l)~V(n-1,\a+\b+1)  : \,\,
		w_0\otimes \left(v_{0,2}-\frac{[n+1]}{[n]}v_{1,1}\right)
        +\frac{[\a+n+1]}{q^{\a+1}[n][\b]}w_1\otimes v_{1,0}\\
        &\phantom{(l)~V(n-1,\a+\b+1)  : \,\,}
        -\frac{[\a+n+1]}{q^{\a+n+1}[\b]}w_2\otimes v_0\,.
	\end{aligned}
	\ee
    Let $\idx = \{s,u,r,l\}$ and $\g=\a+\b$. Then for $x\in\idx$, let $h_x$ and $\widetilde{h}_x$ denote the highest weight vectors given in \eqref{highvexplicit} and  \eqref{highv}, respectively. These are related by some normalization $N_x$.
    \begin{lemma}\label{lem.hwtnormalize}
        For each $x\in\idx$,
        \[
        h_x=\frac{1}{N_x}\widetilde{h}_x\qquad 
        \mbox{where }N_x=
        \begin{cases}
            [\g][\g+n+1] & x=s
            \\
            q^{2\a+n+2}[\b][\b+1]&x=u
            \\
            q^\a[\b][\g+n+2] & x=r
            \\
            q^{\a+n+1}[n][\b][\g+1]&x=l
        \end{cases}\,.
        \]
    \end{lemma}
    \begin{proof}
        We expand the expressions $\widetilde{h}_x$ so that we may compare coefficients. In the case $s=x$, since $w_0\otimes v_0$ is a highest weight vector with highest weight $(n,\g)$, the expansion amounts to computing $E_{0,3}F_{0,3}x_0$ for such a highest weight vector $x_0$. We refer to the actions given in \eqref{eq.action}
        \be
        \begin{aligned}
            \label{eq.x=scase}
            E_2E_1E_2x_{03}&=E_2E_1([\g+1]x_{02}-[\g]x_{1,1})
        =E_2([\g+1][n+1]-[\g][n])x_{0,1}
        \\&=[\g+n+1]E_2x_{0,1}=\underbrace{[\g][\g+n+1]}_{N_s}x_0\,.
        \end{aligned}
        \ee

        For $x=u$, we compare coefficients of $w_3\otimes v_0$. Notice that there is a unique term in the expansion of $\Delta(E_{0,3}F_{0,3})(w_0\otimes F_2F_1F_2v_0)$ which is a multiple of $w_3\otimes v_0$. We write 
        \[
        [w_3\otimes v_0]\Delta(E_{0,3}F_{0,3})(w_0\otimes F_2F_1F_2v_0)
        =
        F_{0,3}w_0\otimes E_{0,3}F_2F_1F_2v_0=[\a][\a+n+1]w_3\otimes v_0\,.
        \]
        Thus 
        \[
        \frac{[\a][\a+n+1]}{N_u}=\frac{[\a][\a+n+1]}{q^{2\a+n+2}[\b][\b+1]}\qquad \mbox{implies} \qquad N_u=q^{2\a+n+2}[\b][\b+1].
        \]

        In the case $x=r$, we use the square nilpotency of $F_2$, which implies $\Delta(E_{0,3}F_{0,3})\Delta(F_2)=0$ and so
        \[
        \Delta(E_{0,3}F_{0,3})(F_2w_0\otimes v_0)=-q^{-\b}\Delta(E_{0,3}F_{0,3})(w_0\otimes F_2v_0).
        \]
        Given that $h_r=w_0\otimes v_{0,1}-q^{-\a}\frac{[\a]}{[\b]}w_1\otimes v_0$ is a highest weight vector of weight $(n+1,\g)$
        \begin{align*}
            [\g][\g+n+2]h_r&=\Delta(E_{0,3}F_{0,3})h_r
            =
            \Delta(E_{0,3}F_{0,3})((1+q^{-\a-\b}\frac{[\a]}{[\b]})w_0\otimes v_{0,1})
            =
            q^{-\a}\frac{[\g]}{[\b]}N_rh_r.
        \end{align*}
        Thus, $N_r=q^\a[\b][\g+n+2]$.
        
        We use a similar approach for $x=l$. Here we use the identity $\Delta(E_{0,3}F_{0,3})\Delta(F_1F_2)=\Delta(E_{0,3}F_{0,3})\Delta(F_2F_1)=0$ which follows from $F_2$ nilpotency and the Serre relation. We consider
        \begin{align*}
            &\Delta([n]F_2F_2-[n+1]F_2F_1)(w_0\otimes v_0)
            \\&\qquad=[n](w_2\otimes v_0 +q^{-1}w_1\otimes v_{1,0}+q^{-\b}w_0\otimes v_{0,2})-[n+1](w_1\otimes v_{1,0}+q^{-\b}w_0\otimes v_{1,1})
            \\&\qquad =
            q^{-\b}w_0\otimes ([n]v_{0,2}-[n+1]v_{1,1})
            +[n]w_2\otimes v_0-q^n w_1\otimes v_{1,0}
        ,
        \end{align*}
        taken together with the previous relation implies
        \begin{align*}
            &\Delta(E_{0,3}F_{0,3})(w_2\otimes v_0 -q^{n}\frac{1}{[n]}w_1\otimes v_{1,0})
        \\&\qquad=
        -q^{-\b}\Delta(E_{0,3}F_{0,3})(w_0\otimes \left(v_{0,2}-\frac{[n+1]}{[n]}v_{1,1}\right))
        .
        \end{align*}
        Since $h_l$ may be written
        \[
        h_l=w_0\otimes \left(v_{0,2}-\frac{[n+1]}{[n]}v_{1,1}\right)
        -\frac{[\a+n+1]}{q^{\a+n+1}[\b]}(w_2\otimes v_0-q^n\frac{1}{[n]} w_1\otimes v_{1,0})
        \]
        applying $\Delta(E_{0,3}F_{0,3})$ gives us
        \begin{align*}
            &\Delta(E_{0,3}F_{0,3})(h_l)
            =
            (1+\frac{[\a+n+1]}{q^{\a+\b+n+1}[\b]}) \Delta(E_{0,3}F_{0,3})(w_0\otimes \left(v_{0,2}-\frac{[n+1]}{[n]}v_{1,1}\right))
            \\&\qquad=
            \frac{[\g+n+1]}{q^{\a+n+1}[\b]}\Delta(E_{0,3}F_{0,3})(w_0\otimes \left(v_{0,2}-\frac{[n+1]}{[n]}v_{1,1}\right))
            = \frac{[\g+n+1]}{q^{\a+n+1}[n][\b]}\widetilde{h}.
        \end{align*}
        Also since $h_l$ is a highest weight vector with weight $(n-1,\g+1)$, 
        \[
        [\g+1][\g+n+1]h_l=\Delta(E_{0,3}F_{0,3})(h_l)=\frac{[\g+n+1]}{q^{\a+n+1}[n][\b]}N_lh_l
        \]
        and so $N_l=q^{\a+n+1}[n][\b][\g+1]$.
    \end{proof}
	
	We assign labels to each summand based on additional shifts of the highest weight in $\BZ_{\geq 0}\times \BC$. Write $V(0,\b)\otimes V(n,\a)=\bigoplus_{x\in\idx} V\left((n,\a+\b)+(\lambda_1,\lambda_2)_x\right)$ where the weight shifts $(\lambda_1,\lambda_2)_i$ are described in Table \ref{tab.shifts} and $x=l$ is omitted if $n=0$.
	\begin{table}[h!]
		\centering
		\[\begin{array}{c|cc}
			i & \lambda_1&\lambda_2  \\
			\hline
			s & 0&0\\
			u & 0&1\\
			r & 1&0\\
			l & -1&1\\
		\end{array}
		\]
		\caption{Weight shifts in the path space.}
		\label{tab.shifts}
	\end{table}
	
	We say that a word $w=x_1\dots x_m\in \idx^m$ is \emph{admissible} if for all left subwords $w[1,j]=x_1\cdots x_j$, the number of $l$'s does not exceed $n$ plus the number of $r$'s. For an admissible word $w$, define 
	\begin{align}\label{eq.pathterms}
	    n_w=n+\sum_{j=1}^m (\lambda_1)_{x_j} \,,
	\qquad
	\a_w = \a +m\b+\sum_{j=1}^m (\lambda_2)_{x_j} \,,
	\qquad \mbox{and} \qquad
	V_w=V(n_w,\a_w).
	\end{align}
    We may also write $n_0=n$, $\a_0=\a$, $V_0=V$ and
    \begin{align}
        n_j=n_{w[1,j]}, \qquad \a_j=\a_{w[1,j]}, \qquad V_j=V_{w[1,j]}.
    \end{align}
    
	The action of $\Uq$ on the highest weight vectors \eqref{highvexplicit} determines the inclusion maps $i_x:V_x\to V(0,\b)\otimes V(n,\a)$.  We define $p_x$ the corresponding splitting of $i_y$ realizing the isomorphism \eqref{V0n}. Thus
	\begin{align}\label{eq.orth}
		p_y\circ i_x = \delta_{xy}\id_{V_x} \qquad \sum_{x\in\idx} i_x\circ p_x=\id_{V(0,\b)\otimes V(n,\a)}
	\end{align}
	We define $i^\dagger_x$ in Section \ref{sec.Herm} which is a multiple of $p_x$ that is compatible with the Hermitian structure. 
	
	For an admissible word $w$ of length $m$, we will use $i_w$ and $p_w$ to denote inclusion and  projection maps into an iterated $m$-fold tensor product $V(0,\b)^{\otimes m}\otimes V(n,\a)$, assuming right-association. If $w=\varnothing$, then we set $V_w=V(n,\a)$, $i_w=\id_{V_w}$, and $p_w=\id_{V_w}$. Now for each admissible word $wx$, let
	\begin{align}
	    i_{x}^{(w)}:V_{wx}\to V(0,\b)\otimes V_w
	\end{align}
	be the appropriate weight shift of $i_x$. We then define $i_{wx}=(\id_{V(0,\b)}\otimes i_w)\circ i_x^{(w)}$, presented as
	\begin{align}
	    V_{wx}\xrightarrow{i_{x}^{(w)}} V(0,\b)\otimes V_w \xrightarrow{\id_{V(0,\b)}\otimes i_w} V(0,\b)^{\otimes (m+1)}\otimes V(n,\a).
	\end{align}
	In particular, all tensor products are right-associated. The projections $p_w$ are defined by the normalization assumption \eqref{eq.orth}.

	\begin{lemma}
		Suppose that $\th$ and $\a$ satisfy \eqref{eq.generic}. Every module obtained as a summand of $V(0,\a)$ under successive fusions with $V(0,-\frac{1}{2})$ has typical highest weight. Thus the stated fusion rule applies.
	\end{lemma}
	\begin{proof}
		After $m$ products of $V(0,-\frac{1}{2})$, the possible weights are of the form $(n,\a-\frac{m}{2}+k)\in \BZ_{\geq 0}\times (\a+\frac{1}{2}\BZ)$. Typicality for all weights requires $[\a+\frac{p}{2}]\neq 0$ for all $p\in \BZ$. Failure of typicality implies $(\a+\frac{p}{2})\th=\pi r$ for some $p,r \in \BZ$, or equivalently after dividing by $\pi$
		\[
		-r+p\cdot \frac{\th}{2\pi}+2\cdot \frac{\a\th}{2\pi}=0,
		\]
		which contradicts the assumption of $\BZ$-linear independence on $\a$ and $\th$.
	\end{proof}
	
	\begin{remark}
		Although $V(0,-\frac{1}{2})\otimes V(0,-\frac{1}{2})$ has non-simple summands associated with atypical weights $(0,-1)$ and $(0,0)$, the tensor products $V(0,-\frac{1}{2})^{\otimes m}\otimes V(0,\a)$ are semisimple under assumption \eqref{eq.generic}. This is most directly realized by decomposing the semisimple right-associated tensor products.
	\end{remark}
	
	\subsection{Braiding and modified trace}
	\label{sub.braiding}
	
	In this section we describe the braiding and modified trace in $\catsl$. We refer to \cite{GHKKW} for further details.
	Recalling the work of Khoroshkin, Tolstoy and Yamane \cite{KhoroshkinTolstoy,Yamane}
	$\Uq$ has an explicit universal quasi-$R$-matrix. Recall $E_{12}$ and $F_{12}$
	from Equation~\eqref{E12F12}. We then define
	\be
	\label{3R}
	\begin{aligned}
		\Check R_1 &= \sum_{i=0}^{\infty} \frac{q^{\tfrac{i(i-1)}{2}}}{[i]!}
		(q-q^{-1})^{i}E_1^i \otimes F_1^i
		\\
		\Check R_2 &= 1\otimes 1 - (q-q^{-1})E_2 \otimes F_2 
		\\
		\Check R_{12} &= 1\otimes 1 - (q-q^{-1})E_{12} \otimes F_{12} 
	\end{aligned}
	\ee
	where $\Check R_1$ belongs to a tensor product completion of $\Uq\otimes \Uq$.
	The quasi-$R$-matrix is then given by $\Check R=\Check R_{2}\Check R_{12}\Check R_{1}$.
	
	For any $V, W \in \catsl$, the action of $\Check R$ on $V\otimes W$ is well-defined
	and we denote this action by the same name $\Check R\in \End_{\BC}(V \otimes W)$.
	Next, define
	\be
	\label{Upsilon}
	\Upsilon_{V,W} \in \End_{\BC}(V \otimes W), \qquad
	\Upsilon_{V,W} (v \otimes w) = q^{\sum(A^{-1})_{ij}\lambda_i \mu_j} v \otimes w
	= q^{-(\lambda_1 \mu_2+\lambda_2 \mu_1+2\lambda_2 \mu_2)} v \otimes w,
	\ee
	where $v\in V$ and $w \in W$ are of weight $\lambda=(\lambda_1,\lambda_2)$ and
	$\mu=(\mu_1,\mu_2)$, respectively and $(A_{ij})$ is the Cartan matrix. For such vectors, we will also write 
    \begin{align}
        \langle \Upsilon_{V,W} (v \otimes w)\rangle =q^{-(\lambda_1 \mu_2+\lambda_2 \mu_1+2\lambda_2 \mu_2)}.
    \end{align}
    Now define the graded swap $\tau_{V,W}$ and the braiding $c_{V,W}$ by
	\be
	\label{cth}
	\begin{aligned}
		\tau_{V,W}: & \,\, V\otimes W\to W\otimes V, \qquad
		\tau_{V,W}(v\otimes w)=(-1)^{\p v \p w}w\otimes v, \\
		c_{V,W} : & \,\, V \otimes W \to W \otimes V, \qquad 
		c_{V,W}=\tau_{V,W}\circ \Upsilon_{V,W}\circ \Check R, \\
        \th_V : & \,\, V \to V, \hspace{2.8cm} \th_V=\trr(c_{V,V})\,.
	\end{aligned}
	\ee
    Recall that the braiding and ribbon structure satisfy the relation
    \begin{align}\label{eq.ribbonbraid}
        \th_{V\otimes W}=(\th_V\otimes \th_W)\circ c_{W,V}\circ c_{V,W}\,.
    \end{align}
	
	The braiding $c_{W,V} \in \Hom_\cat(W\otimes V, V\otimes W)$, where $W=V(0,\b)$ and $V=V(0,\a)$, can be described explicitly using the ordered bases $v_{0,i}$ and $w_{0,i}$  of $V$ and $W$, following the notation of Equation~\eqref{eq.basis}. We further abbreviate the vectors $x_{ij}=w_{0,i}\otimes v_{0,j}$ and $y_{ij}=v_{0,i}\otimes w_{0,j}$ for $i,j=0,\dots,3$. Writing $\g=\a+\b$, it follows that the $(i,j)$-entry of the map $q^{2\a\b}(c_{W,V}(x_{ij}))_{0\leq i,j\leq 3}$ is given by
    \begin{center}
		\small
		\resizebox{\textwidth+2em}{!}{
			$
			\left(\begin{array}{@{}cccc}
				%i=1
				y_{00}
				&
				q^{-\b} y_{10}
				&
				q^{-\b} y_{20}
				&
				q^{-2\b}y_{30}
				\\ %i=2
				q^{-\a} y_{01} + q^{-\b}\br \b y_{10}
				&
				- q^{-\g}y_{11}
				&
				-q^{-\g-1} y_{21}+
				q^{-2\b}\br \b y_{30}
				&
				q^{-\g-\b-1} y_{31}
				\\ %i=3
				q^{-\a} y_{02} + q^{-\b} \br \b y_{20}
				&
				-q^{-\g-1} y_{12}-{q^{-\g-1}\br 1} y_{21}
				-
				q^{-2\b-1}\br \b y_{30}
				&
				- q^{-\g} y_{22}
				&
				q^{-\g-\b-1} y_{32}
				\\ %i=4
				q^{-2\a}y_{03}+q^{-\g-1}\br{\b+1} (y_{12}-q^{-1}y_{21}) + q^{-2\b-1}\br{\b}\br{\b+1} y_{30}
				&
				q^{-\a-\g-1} y_{13}-q^{-\g-\b-1} \br{\b+1} y_{31}
				&
				q^{-\a-\g-1} y_{23}-q^{-\g-\b-1} \br{\b+1} y_{32}
				&
				q^{-2(\g+1)} y_{33}
			\end{array}\right)
			$}\,.
	\end{center}  
	This matrix is computed in \cite{LGTQCGithub}. The ribbon element acts on $V=V(n,\a)$ by the scalar $\th_V=q^{-2\a(n+\a+1)}\id_V$. We also choose a square root $\sqrt\th_V=q^{-\a(n+\a+1)}\id_V$. We denote the scalar action of $\th$ and $\sqrt\th$ by
    \begin{align}
        \label{eq.sqrt}
        \langle\th_V\rangle = q^{-2\a(n+\a+1)}\qquad \mbox{and}\qquad
        \langle\sqrt\th_V\rangle =q^{-\a(n+\a+1)}.
    \end{align}
    
	Our normalization of the modified trace is
	\be\label{eq.mdim}
	\begin{aligned}
		\md(V(n,\a))=(-1)^n\frac{[n+1]}{[\alpha]
			[n+\alpha+1]}\,.
	\end{aligned}
	\ee
	Writing $[x]=\frac{\sin(\th x)}{\sin(\th)}$, we note that for $0<\th<\pi$ the modified dimension of $V(0,\a)$ is positive whenever
	\begin{align}
		\label{eq.mdalpha}
		\frac{k\pi}{\th}<\a<\frac{(k+1)\pi}{\th}-1 \qquad\mbox{for $k\in\BZ$.}
	\end{align}
	In particular, it is positive for $0<\th<\pi$ and $0<\a<\frac{\pi}{\th}-1$.
	
	\section{Hermitian Structure}
	\label{sec.Herm}
	We follow \cite{GLPS22, GLPS24} for the construction of a dagger structure on the quantum group. Let define an anti-superhomomorphism on $\Uq$ by
	\begin{align}\label{eq.daggerH}
		E_1^\dagger=F_1,\qquad 
		E_2^\dagger=-F_2,\qquad 
		H_i^\dagger=H_i,\qquad
		F_1^\dagger=E_1,\qquad 
		F_2^\dagger=E_2,\qquad 
		K_i^\dagger=K_i^{-1},
	\end{align}
	and for $a\in\BC$ and $x,y\in \Uq$
	\begin{align}
	    (ax)^\dagger=\bar{a}x^\dagger
	\qquad (xy)^\dagger=(-1)^{\p x \p y}y^\dagger x^\dagger
	\qquad \Delta(x^\dagger)=\tau\circ \Delta(x)^{(\dagger\otimes \dagger)}
	\end{align}
	where $\tau(x\otimes y)=(-1)^{\p x \p y}(y\otimes x)$. The dagger is a super-involution, i.e.~$(x^\dagger)^\dagger=(-1)^{\p x}x$ for homogeneous $x$. It also satisfies the following Hopf algebraic relations proven in \cite{GLPS24}
	\begin{align}
	    \e(x^\dagger)=\overline{\e(x)},\qquad S(x^\dagger)=S(x)^\dagger,\qquad (R)^{\dagger \otimes \dagger} = \tau(R^{-1})\,.
	\end{align}
	
	A \emph{sesquilinear $\Uq$-module} is a finite dimensnioal $\Uq$-module $V$ equipped with a non-degenerate sesquilinear form $(\cdot|\cdot)_V$ such that for any $c\in \BC$, homogeneous $x\in \Uq$ and $v,w\in V$
	\begin{align}
		(cv\mid w)=\bar c (v\mid w),\qquad  (v\mid  c w)=c(v\mid  w),\quad \mbox{and}\quad
		(xv\mid w)_V=(-1)^{\p x \p v}(v\mid x^\dagger w)_V.
	\end{align}
    
    We recall from \cite[\S II.5.1]{Tu:book} the notion of a dagger structure, or conjugation on a category $\cat$. A \emph{dagger} $\dagger$ is a contravariant involution on $\cat$ which preserves objects and is strictly monoidal. We have for $f\in \Hom_{\cat}(V,W)$, its adjoint $f^\dagger\in \Hom_{\cat}(W,V)$, and for any morphism $g$ between objects in $\cat$,
	\be
	\begin{aligned}
		(f^\dagger)^\dagger=f\qquad
		(f\otimes g)^\dagger=f^\dagger\otimes g^\dagger 
		\qquad 
		(f\circ g)^\dagger=g^\dagger\circ f^\dagger.
	\end{aligned}
	\ee
    We say $\cat$ is a \emph{sesquilinear ribbon category} if the dagger structure has the following properties. For all objects $V,W\in \cat$
\begin{equation}
    \begin{aligned}
       & & c^\dagger_{V, W}&=c_{V, W}^{-1}\,, &
         \th_V^\dagger&=\th_V^{-1}\,, & &\\
         \rcoev_V^\dagger&=\lev_V\,, &
         \lcoev_V^\dagger&=\rev_V\,, &
         \lev_V^\dagger&=\rcoev_V\,, & 
         \rev_V^\dagger&=\lcoev_V\,.
    \end{aligned}
\end{equation}
	Further, $\cat$ is a \emph{Hermitian ribbon category} if, in addition, the dagger is involutive, i.e. 
    \begin{align}
        (f^\dagger)^\dagger=f \qquad\mbox{for every morphism $f$ in $\cat$.}
    \end{align}

    For any two simple objects $V$, $W$ equipped with Hermitian forms, the pairing on $V\otimes W$ is computed in two steps. First, introduce notation for the naive pairing on tensor products. Let $v_1,v_2\in V$ and $w_1,w_2\in W$, then set
	\begin{align}
		\label{eq:naivepairing}
		(v_1\otimes w_1\mid v_2\otimes w_2)^0_{V\otimes W}
		=
		(-1)^{\p {v_2} \p {w_1}}(v_1\mid v_2)_V(w_1\mid w_2)_W.
	\end{align}
	This pairing is not compatible with the quantum group action. To define a pairing which respects tensor products \cite{GLPS22} introduced the notion of a half twist $\sqrt{\th}$ which is a natural isomorphism of the identity functor satisfying 
    \[
    (\sqrt{\th}_V)^2=\th_V \qquad \mbox{and}\qquad (\sqrt{\th}_V)^*=\sqrt\th_{V^*}\,.
    \]
    Define
	\begin{align}
		\label{eq.Xeta}
		\tc_{V,W}=(\sqrt{\th}_{W\otimes V})^{-1}\circ c_{V,W} \circ (\sqrt{\th}_V\otimes \sqrt{\th}_W),
		\qquad 
		\eta_{V,W}=\tau_{W,V}\circ \tc_{V, W}.
	\end{align}
    Our $\tc$ is the map $\sX$ in the notation of \cite[Eq.(3.18)]{GLPS24}.
	Then 
	\begin{align}
		\label{eq.tensorpairing}
		(v_1\otimes w_1\mid v_2\otimes w_2)_{V\otimes W}=
		(v_1\otimes w_1\mid \eta_{V,W}(v_2\otimes w_2))^0_{V\otimes W}
	\end{align}
	is a sesquilinear form compatible with the action of the quantum group on tensor products.
    	
	Recall that $\catslR$ is the full subcategory of $\catsl$  generated by objects with real weights $(n,\a)\in\BZ_{\geq 0}\times \BR$. Let $\catsl_\mathrm{Herm}$ denote the full subcategory of $\catslR$ generated by tensor products of simple Hermitian objects and their duals. We recall the scalar actions $\langle\th_V\rangle = q^{-2\a(n+\a+1)}$ and $\langle\sqrt\th_V\rangle =q^{-\a(n+\a+1)}$ on any simple module $V=V(n,\a)$ given in \eqref{eq.sqrt}.
    
    \begin{proof}[Proof of Theorem \ref{thm.herm}]
        To deduce that the subcategory $\catsl_\mathrm{Herm}$ is Hermitian, we apply \cite[Thm.4.8]{GLPS24}. It suffices to show that $\catslR$ is ribbon sesquilinear, generically semisimple, semisimple tensor products of Hermitian objects are Hermitian, and $\rev_V$ is an epimorphism for each generic simple Hermitian $V(n,\a)$.
        
        Although we do not consider $q$ a root of unity, the definition of the dagger agrees on generators with \cite[Eq.(7.26)]{GLPS24} and so the same verification of \cite[Lem.7.13-7.14]{GLPS24} applies in our setting. Thus, $\catslR$ is ribbon sesquilinear. 

        Next, grade $\catslR$ by $\mathcal{G}=\BR/\BZ$, the $H_2$-weights modulo $\BZ$. Primitivity of $H_2$ implies that this grading is additive under tensor products. Let
        \[
        \mathcal{X}_\th=\left\{\frac{k\pi}{\theta}+\mathbb Z\mid k\in\mathbb Z\right\}
        \]
        be a small and symmetric subset of $\mathcal{G}$. Indeed, $\mathcal{X}$ is small as it is a countable subset of an uncountable group. Every atypical weight has degree in $\mathcal{X}_\th$, so $[\a][n+\a+1]$ is non-vanishing for weights outside of $\mathcal{X}$. As typical weights imply projective by \cite[Lem.3.5]{GHKKW}, any object graded in $\mathcal{G}\setminus\mathcal{X}$ is semisimple. This establishes generic semisimplicity.

        The real-highest-weight argument of \cite[Lem.7.16 \& 3.11]{GLPS24} equips each simple object with a Hermitian form. Hermitian symmetry and nondegeneracy imply adjoints on morphisms between these objects is involutive. Moreover, semisimple tensor products of Hermitian simple objects are again Hermitian by \cite[Proposition 3.13]{GLPS24}. As $\rev_V$ is a nonzero map to the tensor unit, it is  indeed an epimorphism for every generic simple Hermitian object $V$.
    \end{proof}

	Since $H_i^\dagger=H_i$, the pairing $(v_{ij}\mid v_{kl})_{V_p(n,\a)}$ is nonzero only if the vectors have equal weight. We give an explicit description of the pairing on Kac modules below.
    
	\begin{proposition}\label{prop.form}
		Let $(n,\a)\in\BZ_{\geq0}\times \BR$ be a typical weight and $V=V_p(n,\a)$. There exists a unique sesquilinear form $(\cdot|\cdot)_V$ such that $(v_0\mid v_0)_V=1$ defined as follows:
		\begin{align*}
		    (v_{ij}\mid v_{ij})_V&=
		\qbin{n}{i}([i]!)^2\cdot
		\begin{cases}
			1&j=0
			\\
			(-1)^p[\a+i]&j=1
			\\
			(-1)^p([n-2i+1][\a+i]+[i][n-i+1][\a+i-1]) & j=2
			\\
			-[\a][\a+n+1] &j=3
		\end{cases}\,,
        \\
        (v_{i,2}\mid v_{i+1,1})_V
		&=
		(v_{i+1,1}\mid v_{i,2})_V
		=
		(-1)^p[\a+i][i+1][n-i]\qbin{n}{i}([i]!)^2,
		\end{align*}
        and it is zero on all other pairs of basis vectors.
	\end{proposition}
	\begin{proof}
		These follow inductively from the relations 
		\[
		(F_1 v \mid w)=(v\mid E_1w)
		\qquad \mbox{and}\qquad
		(F_2 v\mid w) = (-1)^{\p v}(v\mid E_2 w)
		\]
		together with the actions $E_kv_{ij}$ given in \eqref{eq.action}. It suffices to consider pairs of vectors with equal weight. Invertiblilty of the Cartan matrix implies that the multiplicities between the simple lowering operators are equal. Thus, we deduce the off-diagonal pair $v_{i+1,1}$ and $v_{i,2}$.
		
		We compute for $i\geq 1$,
		\begin{align*}
		    (v_{i,0}\mid v_{i,0})= (F_1v_{i-1,0}\mid v_{i,0})=(v_{i-1,0}\mid E_1v_{i,0})
		=[i][n-i+1](v_{i-1,0}\mid v_{i-1,0})=\prod_{j=1}^i[j][n-j+1]\,.
		\end{align*}
        We will use the notation 
        \[
        B_i=\prod_{j=1}^i[j][n-j+1]=\qbin{n}{i}([i]!)^2=	(v_{i,0}\mid v_{i,0}) \qquad \mbox{with $B_0=1$}.
        \]
		The computations of the pairings $(v_{i,1}\mid v_{i,1})$ and $(v_{i,2}\mid v_{i+1,1})$ are also straightforward:
        \begin{align*}
            (v_{i,1}\mid v_{i,1})&=
		(F_2v_{i,0}\mid v_{i,1})=
		(-1)^p(v_{i,0}\mid E_2 v_{i,1})=
		(-1)^p[\a+i](v_{i,0}\mid v_{i,0})\\&
        \qquad=
		(-1)^p[\a+i]B_i \,,
        \\
        (v_{i,2}\mid v_{i+1,1})&=
		(F_1v_{i,1}\mid v_{i+1,1})=
		(v_{i,1}\mid E_1v_{i+1,1})=
		[i+1][n-i](v_{i,1}\mid v_{i,1})
        \\&\qquad=
		[i+1][n-i]\cdot (-1)^p[\a+i]B_i\,.
        \end{align*}
		This particular pairing is symmetric because $(v_{i+1,1}\mid F_1v_{i,1})=(E_1v_{i+1,1}\mid v_{i,1})$ again leads to the symmetric pairing of $v_{i,1}$ with itself. Next, consider $(v_{i,2}\mid v_{i,2})$ worked out below.
		\begin{align*}
			(v_{i,2}\mid v_{i,2})&=
			(F_1v_{i,1}\mid v_{i,2})=
			(v_{i,1}\mid E_1v_{i,2})
			\\&
			=
			[n-2i+1](v_{i,1}\mid v_{i,1})
			+[i][n-i+1](v_{i,1}\mid v_{i-1,2})
			\\&
			=
			[n-2i+1]\cdot 	(-1)^p[\a+i]B_i
			+[i][n-i+1]\cdot (-1)^p[\a+i-1][i][n-i+1]B_{i-1}
			\\&
			=
			(-1)^pB_{i}([i][n-i+1][\a+i-1]+[n-2i+1][\a+i]).
		\end{align*}
		For the remaining calculation of $(v_{i,3}\mid v_{i,3})$, we provide a separate induction on $i$. First, the base case is equivalent to the calculation \eqref{eq.x=scase}
        \begin{align*}
            (v_{03}\mid v_{03})&=(F_2F_1F_2 v_0\mid v_{03})
            =-(v_0\mid E_2E_1E_2v_{03})
            =-[\a][\a+n+1](v_0\mid v_0)=-[\a][\a+n+1].
        \end{align*}
	Then inductively, for $i\geq 1$
    \begin{align*}
        (v_{i,3}\mid v_{i,3})&=(F_1v_{i-1,3}\mid v_{i,3})=(v_{i-1,3}\mid E_1v_{i,3})
        \\
        &=[i][n-i+1](v_{i-1,3}\mid v_{i-1,3})
        \\
        &=B_i(v_{03}\mid v_{03})=-B_i[\a+n+1][\a].\qedhere
    \end{align*}
	\end{proof}

    For simple projective $V\in\catsl_\mathrm{Herm}$ and any $W\in\catsl_\mathrm{Herm}$, define a pairing on $\Hom_{\catsl_\mathrm{Herm}}(V,W)$ given by
    \begin{align}\label{eq.hompairing}
        \langle f,g\rangle = \mt_V(f^\dagger \circ g)=\md(V)(v_0\mid f^\dagger \circ g (v_0))_V = \md(V) (f(v_0)\mid g(v_0))_W
    \end{align}
	where $v_0\in V$ is a highest weight vector and $(v_0\mid v_0)_V=1$.
	We are now prepared to compute $i_x^\dagger = \nu_x p_x$ as a rescaling of the projector $p_x$.
	
	\begin{proposition}
    \label{prop.nu}
		For each $i_x$ whose target is $V(0,\b)\otimes V(n,\a)$ and $x\in\{s,u,r,l\}$, 
		\[
		\nu_x(n,\a;\b)=
		\begin{cases}
			1 &x=s
			\\[2ex]
			-\dfrac{[\a][\a+n+1][\a+\b+1][\a+\b+n+2]}{[\b][\b+1]} & x=u
			\\[2ex]
			(-1)^n\dfrac{[\a][\a+\b]}{[\b]} & x=r
			\\[2ex]
			(-1)^n\dfrac{[n+1]}{[n]}\dfrac{[\a+n+1][\a+\b+n+1]}{[\b]} & \mbox{$x=l$ and $n\geq 1$}
		\end{cases}.
		\]
	\end{proposition}
	\begin{proof}
		For $x\in \{s,u,r,l\}$, let $h_x=i_x(v_0)\in W\otimes V$ as in \eqref{highvexplicit} with $W=V(0,\b)$, $V=V(n,\a)$, and $V_x$ as in \eqref{eq.pathterms}. To determine $\nu_x$, first observe that
        \[
        (h_x\mid h_x)_{W\otimes V} = (i_x(v_0)\mid i_x(v_0))_{W\otimes V}
        =
        (v_0\mid i_x^\dagger\circ i_x(v_0))=\nu_x(v_0\mid j_x\circ i_x(v_0))=\nu_x.
        \]
        Thus, it suffices to compute 
        \[
        (h_x\mid h_x)_{W\otimes V}=(h_x\mid \eta_{V,W}h_x)^0_{V\otimes W}
        =
        \frac{\langle \th_{V}^{1/2}\rangle \langle \th_{W}^{1/2}\rangle}{\langle \th_{V_x}^{1/2}\rangle}
        (h_x\mid \tau_{W,V}\circ c_{V,W} h_x)^0_{V\otimes W}
        \]
        where $\langle \th_{V}^{1/2}\rangle$ is the half-twist scalar in \eqref{eq.sqrt}. For each $x$, write $v_x$ for the right tensor factor acted by $\Delta(E_{0,3}F_{0,3})$ in \eqref{highv}. Now we simplify $\tau_{W,V}\circ c_{V,W} h_x$ as follows, where we recall that $h_x=N_x^{-1}\Delta(E_{0,3}F_{0,3})(w_0\otimes v_x)$, the braiding commutes with $\Delta(E_{0,3}F_{0,3})$, and the quasi-$R$-matrix acts by the identity on $w_0\otimes v_x$:
        \[
        \begin{aligned}
            \tau_{W,V}\circ c_{V,W} h_x&=N_x^{-1}\tau_{W,V}\circ c_{V,W} \Delta(E_{0,3}F_{0,3})(w_0\otimes v_x)
            \\
            &=N_x^{-1}\tau_{W,V}\Delta(E_{0,3}F_{0,3}) c_{V,W}(w_0\otimes v_x)
            \\
            &=
          N_x^{-1}\langle \Upsilon(w_0\otimes v_x) \rangle \Delta^{\mathrm{op}}(E_{0,3}F_{0,3})(w_0\otimes v_x),
        \end{aligned}
        \]
        where $\langle \Upsilon(w_0\otimes v_x)\rangle =q^{-\b(\mu_1-2\mu_2)}$ is the scalar part of the diagonal action and $v_x$ has weight $(\mu_1,\mu_2)$. We now pass the action through the pairing, and apply the identities $\Delta(E_{0,3}F_{0,3})^{\dagger\otimes \dagger}=\Delta^{\mathrm{op}}((E_{0,3}F_{0,3})^\dagger)$ and $(E_{0,3}F_{0,3})^\dagger=E_{0,3}F_{0,3}$:
        \[
        (h_x\mid \Delta^{\mathrm{op}}(E_{0,3}F_{0,3})(w_0\otimes v_x))^0
        =
        {\overline {N_x}^{-1}}(w_0\otimes v_x\mid \Delta^{\mathrm{op}}(E_{0,3}F_{0,3})\cdot \Delta^{\mathrm{op}}(E_{0,3}F_{0,3})(w_0\otimes v_x))^0.
        \]
        Since $\Delta^{\mathrm{op}}(E_{0,3}F_{0,3})(w_0\otimes v_x))$ is also a highest weight vector of weight $(n_x,\a_x)$ with respect to the opposite action, the action of $\Delta^{\mathrm{op}}(E_{0,3}F_{0,3})$ on it agrees with the scalar $[\a_x][\a_x+n_x+1]$ computed in the $(i,3)$ part of the proof of Proposition \ref{prop.form} without the grading sign.

        Before collecting all of the terms, we calculate
        \[
        (w_0\otimes v_x\mid \Delta^{\mathrm{op}}(E_{0,3}F_{0,3})(w_0\otimes v_x))^0
        =(\Delta(E_{0,3}F_{0,3})w_0\otimes v_x\mid w_0\otimes v_x)^0
        =\overline{N_x}(h_x\mid w_0\otimes v_x)^0
        \]
        which amounts to pairing the right tensor factor in $h_x$ with $v_x$. In each case, we have
        \[
        (h_x\mid w_0\otimes v_x)^0=
        \begin{cases}
            (v_0\mid v_0)_V=1 & x=s
            \\
            (v_{0,3}\mid v_{0,3})_V=-[\a][\a+n+1] &x=u
            \\
            (v_{0,1}\mid v_{0,1})_V=(-1)^n[\a] & x=r
            \\
            (v_{0,2}-\frac{[n+1]}{[n]}v_{1,1}\mid [n]v_{0,2}-[n+1]v_{1,1})_V & x=l
        \end{cases}
        \]
        where the pairing  $(v_{0,2}-\frac{[n+1]}{[n]}v_{1,1}\mid [n]v_{0,2}-[n+1]v_{1,1})_V$ in the $x=l$ case further simplifies to
        \begin{align*}
            &[n](v_{0,2}\mid v_{0,2})_V
        -2[n+1](v_{0,2}\mid v_{1,1})_V+
        \frac{[n+1]^2}{[n]}(v_{1,1}\mid v_{1,1})_V
        \\
        &\qquad=(-1)^n[n][n+1][\a]
        -2(-1)^n[n+1][\a][n]
        +(-1)^n\frac{[n+1]^2}{[n]}[n][\a+1]
        \\&\qquad=(-1)^n[n+1]([n+1][\a+1]-[n][\a])
        =(-1)^n[n+1][n+\a+1].
        \end{align*}
        Therefore, using the values in \eqref{lem.hwtnormalize}, \eqref{eq.sqrt}, and the pairing above, we have $\nu_x$ given by
        \begin{align*}
        &\frac{\langle \th_{V}^{1/2}\rangle \langle \th_{W}^{1/2}\rangle}{\langle \th_{V_x}^{1/2}\rangle}N_x^{-1}\langle \Upsilon(w_0\otimes v_x)\rangle [\a_x][\a_x+n_x+1](h_x\mid w_0\otimes v_x)^0.
        \end{align*}
        The $q$-exponent given by  $\dfrac{\langle \th_{V}^{1/2}\rangle\langle \th_{W}^{1/2}\rangle}{\langle \th_{V_x}^{1/2}\rangle}\langle \Upsilon(w_0\otimes v_x)\rangle$ is
        \begin{align*}
            &q^{-\a(n+\a+1)-\b(\b+1)+\a_x(n_x+\a_x+1)-\b(\mu_1+2\mu_2)}
            \\&\qquad=
            \begin{cases}
                q^{-\a(n+\a+1)-\b(\b+1)+(\a+\b)(n+\a+\b+1)-\b(n+2\a)} &x=s
              \\
              q^{-\a(n+\a+1)-\b(\b+1)+(\a+\b+1)(n+\a+\b+2)-\b(n+2(\a+1))}&x=u
                \\
               q^{-\a(n+\a+1)-\b(\b+1)+(\a+\b)(n+\a+\b+2)-\b(n+1+2\a)} &x=r
               \\
               q^{-\a(n+\a+1)-\b(\b+1)+(\a+\b+1)(n+\a+\b+1)-\b(n+2\a+1)}  &x=l
            \end{cases}
            \quad=
            \begin{cases}
                1 &x=s\\
              q^{2\a+n+2}&x=u
                \\
               q^{\a} &x=r
               \\
               q^{\a+n+1}  &x=l
            \end{cases}\,.
        \end{align*}
       Which gives
            \begin{align*}
            \nu_x&=q^{-\a(n+\a+1)-\b(\b+1)+\a_x(n_x+\a_x+1)-\b(\mu_1+2\mu_2)}[\a_x][\a_x+n_x+1]\frac{(h_x\mid w_0\otimes v_x)^0}{N_x}
            \\&=
            \begin{cases}
              1\dfrac{[\g][\g+n+1]}{[\g][\g+n+1]}  &x=s
              \\[2ex]
              q^{2\a+n+2}
              \dfrac{-[\g+1][\g+n+2][\a][\a+n+1]}{q^{2\a+n+2}[\b][\b+1]} &x=u
                \\[2ex]
                q^{\a} 
               \dfrac{(-1)^n[\g][\g+n+2][\a]}{q^\a[\b][\g+n+2]} &x=r
               \\[2ex]
               q^{\a+n+1}
               \dfrac{(-1)^n[\g+1][\g+n+1][n+1][n+\a+1]}{q^{\a+n+1}[n][\b][\g+1]} &x=l
            \end{cases}
        \end{align*}
        and simplifies to the desired expressions.
	\end{proof}
	
	\begin{proposition}\label{prop.pathnu}
		Let $w$ be an admissible fusion path in $V(0,\b)^{\otimes m}\otimes V(n,\a)$. Then
        \[
        i_w^\dagger = \nu_w p_w \qquad
        \mbox{where} \qquad  \nu_w=\prod_{j=1}^m\nu_{x_j}(n_{j-1},\a_{j-1};\b)
        \]
        following the notation of \eqref{eq.pathterms}. In particular, for paths $w$, $w'$ satisfying $V_w=V_{w'}$, then \[\langle i_w,i_{w'}\rangle = \delta_{w,w'}\nu_w\md(V_w).\]
	\end{proposition}
    \begin{proof}
        For the empty path $\nu_\varnothing=1$. We proceed by induction on the path length with the dagger of the inclusion map satisfying
        \[
        i_{wx}^\dagger=((\id \otimes i_w)\circ i_x^{(w)})^\dagger
        =\nu_w \nu_x^{(w)} j_x^{(w)} (\id\otimes j_w)=\nu_{wx}j_{wx},
        \]
        where $\nu_x^{(w)}=\nu_{x}(n_{j-1},\a_{j-1};\b)$ is the appropriate shift of $\nu_{x}$. The latter claim follows from orthonormality between $i_w$ and $p_w'$, and the definition of the pairing \eqref{eq.hompairing}.
    \end{proof}

	\section{Topological quantum computation}\label{sec.tqc}
	We use an encoding similar to \cite{IKLS25,IKLS26} for our qubits. The key difference is that our tensor product is right associated, which simplified our calculation of $\nu$ in Proposition \ref{prop.nu}. It is also worth noting that our construction consists of only neglectons. For $\a$ and $\th$ satisfying the arithmetic constraint \eqref{eq.generic}, we set 
    \be
	\mbox{$\bW=V(0,-\frac{1}{2})$ and $\bV=V(0,\a)$.}
	\ee
    and for each $N\geq 0$,
	\begin{equation}
		\begin{aligned}
			\hilb_N=\Hom(\bV,  \bW^{2N}\otimes \bV)
			\qquad \mbox{and}
			\qquad
			\hilb^{\mathrm{comp}}_N=\mathrm{span}\{i_w\mid w\in\{su,us\}^N\}\subset \hilb_N.
		\end{aligned}
	\end{equation}
	where the total Hermitian space $\hilb_N$ is considered in the right-associated tensor product and $\hilb^{\mathrm{comp}}$, our computational space, is the $2^N$-dimensional space indexed by fusion paths of length $2N$. We consider $N=1$ and $N=2$ as explicit examples. We have three morphisms in the total space	
	\[
	i_{su},i_{us},i_{rl}\in \Hom_{\catsl_\mathrm{Herm}}(\bV,\bW\otimes \bW\otimes \bV)=\hilb_1
	\]
	and we write $\ket{0}=i_{su}$ and $\ket{1}=i_{us}$ for the generators of our single qubit computational Hilbert space $\hilb_1^{\mathrm{comp}}$, the $i_{rl}$ channel is leakage. The set of labels for the 20-dimensional $\hilb_2$ are exactly the length four words:
	\begin{align*}
		\underline{usus},\underline{ussu},usrl,
		\underline{suus},\underline{susu},surl,
		rlus,rlsu,rlrl,
		uuss,\\
		ursl,ssuu,
		srul,ruls,rslu,
		rrll,rusl, rsul,
		srlu,urls
	\end{align*}
	and the four underlined words form the computational basis $\ket{11}$, $\ket{10}$, $\ket{01}$, $\ket{00}$.

    Computing the matrix of inner products $M_1$ of the $i_w\in\hilb_1$ defines the inner product form in our basis. Proposition \ref{prop.pathnu} tells us that this pairing matrix is diagonal and computed from $\nu_x$ in Proposition \ref{prop.nu}. 
    \begin{align}
        M_1&=\md(\bV)\begin{pmatrix}
            \nu_s(0,\a;-\frac{1}{2})\nu_u(0,\a-\frac{1}{2};-\frac{1}{2}) & 0 & 0
            \\
            0&\nu_u(0,\a;-\frac{1}{2})\nu_s(0,\a+\frac{1}{2};-\frac{1}{2}) & 0
            \\
            0 & 0 &\nu_r(0,\a;-\frac{1}{2})\nu_l(1,\a-\frac{1}{2};-\frac{1}{2})
        \end{pmatrix} 
        \\
        &=\md(\bV)\begin{pmatrix}
            \dfrac{[\a-\frac{1}{2}][\a][\a+\frac{1}{2}][\a+1]}{[\frac{1}{2}]^2} & 0 & 0
            \\
            0&\dfrac{[\a][\a+\frac{1}{2}][\a+1][\a+\frac{3}{2}]}{[\frac{1}{2}]^2} & 0
            \\
            0 & 0 &-\dfrac{[\a-\frac{1}{2}][\a][\a+1][\a+\frac{3}{2}][2]}{[\frac{1}{2}]^2}
        \end{pmatrix} 
    \end{align}
    Identifying $[x]=\frac{\sin(x\th)}{\sin(\th)}$, we see that for $(\th,\a)\in \mathcal{R}$, where
    \begin{align}
        \label{eq.region}
        \mathcal{R}=\left\{(\th,\a)\in \BR^2\mid
        0<\th<\frac{\pi}{2},\quad \frac{1}{2}<\a<\frac{\pi}{\th}-\frac{3}{2}\right\},
    \end{align}
    all the entries $[\a-\frac{1}{2}]$, $[\a]$, $[\a+\frac{1}{2}]$, $[\a+1]$, $[\a+\frac{3}{2}]$, and $[2]$ are positive. The modified dimension given in \eqref{eq.mdim} is also positive in this region. Thus, $M_1$ has signature $(2,1)$ for such $\th$ and $\a$.

    For $f=\sum f_x i_x$ and $g=\sum g_x i_x$, $\langle f, g\rangle = f^* M_1 g$. Further, we write 
    \begin{align}
        \mbox{$M_1=D_1^*\Sigma_1 D_1$	where $D_1=|M_1|^{1/2}$ and $\Sigma_1=\mathrm{diag}(1,1,-1)$.}
    \end{align}
    Recall that for an $n\times n$ nondegenerate Hermitian matrix $H$, the unitary and special unitary operators with respect to the form $H$ are
	\begin{align}
	    \mathrm{U}(H)=\{A\in\mathrm{GL}_n(\BC)\mid A^*HA=H\}\quad\mbox{and}\quad 
	\mathrm{SU}(H)=\{A\in \mathrm{U}(H)\mid \mathrm{det}(A)=1\}.
	\end{align}
    If $H$ is a diagonal matrix with $s_+$ $1$'s and $s_-=n-s_+$ $-1$'s we write $\mathrm{U}(s_+,s_-)$ and $\mathrm{SU}(s_+,s_-)$, respectively. 
    In the case of $H=M_1$, the map $A\mapsto D_1AD^{-1}_1$ establishes an isomorphism between $\mathrm{U}(M_1)$ and $\mathrm{U}(\Sigma_1)=\mathrm{U}(2,1)$.

	Write $\si_i$ for the $i$-th standard generator of the braid group $B_n=B_{2N+1}$. We use the notation 
	\begin{align}
		\label{eq.JM}
		J_{i}=\si_{i}\si_{i+1}\cdots \si_{n-2}\si_{n-1}^2\si_{n-2}\cdots \si_{i+1}\si_{i} \qquad 1\leq i \leq n-1
	\end{align}
	to denote the $i$-th (right-side) Jucys-Murphy element. For $i=n-1$, $J_{n-1}=\si_{n-1}^2$. The set of $J_i$ generate an abelian subgroup of the pure braid group. 

    Let $B_{2N,1}$ be the group of braids which preserve the $2N+1$-st component. We denote the action of $B_{2N,1}$ on the $\hilb_N$ via the categorical braiding by $\rho_N$. Such an action fixes the last component $\bV$ in the definition of $\hilb_N$ so that $\rho_N(B_{2N,1})\subseteq \End(\hilb_N)$.

    Since $M_1$ is associated to a Hermitian pairing in $\catsl_{\mathrm{Herm}}$ and the braiding satisfies $c_{V,W}^{\dagger}=c_{V,W}^{-1}$, it is an isometry $(c_{V,W}(v_1\otimes w_1)\mid c_{V,W}(v_2\otimes w_2))=(v_1\otimes w_1\mid v_2\otimes w_2)$. In particular, the relation $\rho_1(b)^* M_1 \rho_1(b)=M_1$ for $b\in B_{2,1}$ implies the image of $\rho_1$ belongs to $\mathrm{U}(M_1)$. 
	
	The actions of Jucys-Murphy elements on the full Hermitian space $\hilb_N$ act diagonally because of naturality with the inclusion maps. This action is given for $J_{n-1}=J_{2N}$ in terms of ribbon actions using \eqref{eq.ribbonbraid},
    \begin{align}
        \th_{W\otimes V}\circ i_x=(\th_W\otimes \th_V)\circ c_{V,W}\circ c_{W,V}\circ i_x \qquad \mbox{implies} \qquad
        c_{V,W}\circ c_{W,V}\circ i_x = \frac{\langle \th_{V_x}\rangle}{\langle \th_{\bV}\rangle\langle \th_{\bW}\rangle}i_x
    \end{align}
    where $x\in\idx$. Thus, $\rho_1(J_{2})\circ i_{xy}=q^{2\a(\a+1)-2\a_x(n_x+\a_x+1)-1/2}\circ i_{xy}$ for $xy\in \{su,us,rl\}$. This action is expressed in the given basis as
	\begin{align}
		\label{eq.J1}
		\rho_1(J_{2})=\mathrm{diag}(q^{2\a},q^{-2(\a+1)},q)=q^{\a X+Y}, \quad \mbox{with $X=\mathrm{diag}(2,-2,0)$ and $Y=\mathrm{diag}(0,-2,1)$.}
	\end{align}
	
	\subsection{Single qubit universality} We prove density of the braid group action in $\mathrm{U}(M_1)$ acting on $\hilb_1$. We begin with a key lemma which states sufficient conditions for the Lie algebra generated by a pair of complex matrices to contain a special linear Lie algebra. We then extend this results to the corresponding Lie groups and apply this to the subgroup generated by $\si_1$ and $J_2$ to prove single qubit universality. 
	
	\begin{lemma}\label{lem.sl}
		Let $\mathfrak{k}\subseteq \mathfrak{gl}(V)$ be a complex Lie subalgebra, where $V=\bigoplus_{s=1}^r V_s \oplus L$. Fix basis $\{e_i\}$ of $V$ adapted to this decomposition. Suppose that there exist $C,D\in\mathfrak{k}$ such that:
		\begin{itemize}
			\item $D=\mathrm{diag}(d_i)$ is diagonal, $C$ preserves the components of $V$, and $[D,C]\vert_L=0$,
			\item the sets of differences $\Delta_s=\{d_i-d_j\mid e_i,e_j\in V_s\}$, satisfy the properties 
            \[
            |\Delta_s|=2{\binom{\dim V_s}{2}}+1,\qquad \mbox{and}\qquad \mbox{$\Delta_s\cap\Delta_t=\varnothing$ for $s\neq t$,}
            \]
			\item each pair $(D\vert_{V_s},C\vert_{V_s})$ acts irreducibly on $V_s$.
		\end{itemize}
		Then $\bigoplus_{s=1}^r \mathfrak{sl}(V_s)\oplus 0_L\subseteq\mathfrak{k}$. 
	\end{lemma}
	\begin{proof}
    The assumptions on $\Delta_s$ imply that within $V_s$ all $D$-eigenvalues are distinct and the differences of eigenvalues within each block are distinct across all blocks.
    
	Let $E_{ij}$ denote the $(i,j)$ matrix unit on $\bigoplus_{s=1}^r V_s$. We have $[D,E_{ij}]=(d_i-d_j)E_{ij}$, so that $E_{ij}$ is an eigenvector of $\mathrm{ad}_D$. Write $\lambda_{ij}=d_i-d_j$, then set 
		\[
		\Omega=\bigcup_{s=1}^r \Delta_s
		\qquad \mbox{and}\qquad
		P_{ij}(x)=\prod_{\mu\in \Omega\setminus\{\lambda_{ij}\}}
		\frac{x-\mu}{\lambda_{ij}-\mu}~\mbox{for $i\neq j$}.
		\]
		Note that $0\in \Omega$ implies $P_{ij}(0)=0$. Distinctness of the $\lambda_{ij}\neq 0$ on $V$ imply $P_{ij}$ is well-defined. Moreover, this implies $P_{ij}(\lambda_{kl})=\delta_{ik}\delta_{jl}$. Now, writing $C=\sum_{kl} c_{kl}E_{kl}+C_L$,
		\[
		P_{ij}(\mathrm{ad}_D)(C)=\sum_{kl}c_{kl} P_{ij}(\mathrm{ad}_D)(E_{kl})+P_{ij}(\mathrm{ad}_D)(C_L)=\sum_{kl}c_{kl} P_{ij}(\lambda_{kl})E_{kl}+P_{ij}(0)C_L=c_{ij}E_{ij}+0.
		\]
		Thus, for $i\neq j$ and $c_{ij}\neq 0$, $E_{ij}\in\mathfrak{k}$.
        
        Now we show that off-diagonal matrix units $E_{ij}$ for which $c_{ij}=0$ belong to $\mathfrak{k}$. Fix $j$ associated to a basis vector index in $V_s$. Let
        \[
        I_j=\{j\}\cup\{k\neq j\mid v_k\in V_s,\,E_{kj}\in\mathfrak{k}\} \qquad \mbox{and}\qquad 
        W_j=\mathrm{span}\{e_{k}\mid k\in I_j\}.
        \]
        Suppose for the sake of a contradiction that $W_j\subsetneq V_s$. Since $W_j$ is $D$-invariant, irreducibility of $(D\vert_{V_s},C\vert_{V_s})$ implies that $W_j$ is not $C$-invariant. Therefore, there exists a pair $a\in I_j$ and $i\notin I_j$ such that $c_{ia}\neq 0$. Thus, $E_{ia}\in \mathfrak{k}$. Since $a\in I_j$, we have $E_{aj}\in \mathfrak{k}$. If $a=i$, then $E_{ij}=E_{aj}\in \mathfrak{k}$, or $a\neq i$ and $[E_{ia},E_{aj}]=E_{ij}\in \mathfrak{k}$. Thus, reaching a contradiction, we have $W_j=V_s$ and all off-diagonal matrix units belong to $\mathfrak{k}$. Moreover, the relation $[E_{ij},E_{ji}]=E_{ii}-E_{jj}$, exhibits all traceless diagonal matrices. Thus producing a set of generators for each $\mathfrak{sl}(V_s)$ in $\mathfrak{k}$.
	\end{proof}

	\begin{corollary}\label{cor.su}
		Let $H$ be a nondegenerate Hermitian matrix which respects the decomposition $V=\bigoplus_{s=1}^r V_s \oplus L$ and write $H_s=H\vert_{V_s}$ . Let $K\leq \mathrm{U}(H)$ be a closed subgroup with $\mathfrak{k}$ its Lie algebra, and $D,C\in\mathfrak{k}_\BC$, the complexification of $\mathfrak{k}$, satisfying the assumptions of Lemma \ref{lem.sl}. Then $\prod_{s=1}^r\mathrm{SU}(H_s)\times \{1_L\}\subseteq K$.
	\end{corollary}
	\begin{proof}
		The previous lemma shows that $\mathfrak{sl}(V_s)$ is contained in $\mathfrak{k}_{\BC}$. 
        Recall that 
        \[\mathfrak{u}(H)=\{A\in\mathfrak{gl}(V)\mid HA+A^*H=0\},\qquad 
        \mathfrak{su}(H)=\{A\in\mathfrak{u}(H)\mid \mathrm{tr}{A}=0\}.
        \]
        Now
        \[
        \mathfrak{su}(H_s)=\mathfrak{sl}(V_s)\cap \mathfrak{su}(H)\subseteq
        \mathfrak{k}_\BC\cap \mathfrak{u}(H)=\mathfrak{k}.
        \]
        Connectedness of $\mathrm{SU}(H_k)$ implies that elements of $\mathrm{SU}(H_k)$ are products of exponentials of $\mathfrak{su}(H_k)$ \cite[Cor.2.31]{Hall}. Thus, $\mathrm{SU}(H_k)\times \{1_L\}\subseteq K$. 
	\end{proof}
	Let us compute the action of $\si_1$ with respect to the right-associated matrix units $i_x$ relative to the ordered basis $x\in\{su,us,rl\}$. More precisely, the $(y,x)$-entry of $\si_1$ is given by $j_y\circ(\id\otimes c_{X,X})\circ i_x$:
	\begin{align}
		\label{eq.sigma2}
		\rho_1(\si_1)=\begin{pmatrix}
			-\dfrac{\br{\frac{1}{2}}^2}{q^{\a}\br{\a-\frac{1}{2}}\br{\a+\frac{1}{2}}} 
			&
            \dfrac{q^{\a}\br{\a}\br{\a+1}}{\br{\a-\frac{1}{2}}\br{\a+\frac{1}{2}}} 
			&
            \dfrac{[2]\br{\frac{1}{2}}\br{\a}}{q\br{\a-\frac{1}{2}}\br{\a+\frac{1}{2}}}
			\\[2ex]
			\dfrac{q^{\a+2}\br{\a}\br{\a+1}}{\br{\a+\frac{1}{2}}\br{\a+\frac{3}{2}}} 
			&
            -
            \dfrac{q^{3\a+2}\br{\frac{1}{2}}^2}{\br{\a+\frac{1}{2}}\br{\a+\frac{3}{2}}} 
			&
            -
            \dfrac{q^{2\a+1}[2]\br{\frac{1}{2}}\br{\a+1}}{\br{\a+\frac{1}{2}}\br{\a+\frac{3}{2}}}
			\\[2ex]
			-
            \dfrac{\br{\frac{1}{2}}\br{\a}}{\br{\a-\frac{1}{2}}\br{\a+\frac{3}{2}}} 
            &
            \dfrac{q^{2\a}\br{\frac{1}{2}}\br{\a+1}}{\br{\a-\frac{1}{2}}\br{\a+\frac{3}{2}}} 
			&
            \left(q^{\a-1}+\dfrac{q^{\a-1}[2]\br{\frac{1}{2}}^2}{\br{\a-\frac{1}{2}}\br{\a+\frac{3}{2}}}\right)
		\end{pmatrix}\,.
	\end{align}
    This was computed using the code available at \cite{LGTQCGithub}. 
    
	\begin{theorem}\label{thm.onequbit}
		Suppose $(\th,\a)\in\mathcal{R}$ satisfies the $\BZ$-independence condition \eqref{eq.generic}. Then the group action $\langle \rho_1(\si_1), \rho_1(J_2)\rangle$ is dense in $\mathrm{U}(M_1)$.
	\end{theorem}
	\begin{proof}
		Let $K=\overline{\langle \rho_1(\si_1), \rho_1(J_2)\rangle}$ be the closure of the braid group action generated by $\rho_1(\si_1)$ and $\rho_1(J_2)$. Let $\mathfrak{k}$ be its Lie algebra.
		
		Recall $\rho_1(J_2)=q^{\a X+Y}=e^{i\th(\a X+Y)}$, where $X=\mathrm{diag}(2,-2,0)$ and $Y=\mathrm{diag}(0,-2,1)$ as in \eqref{eq.J1}. The $\BZ$-linear independence assumption and Kronecker's Theorem imply that the subgroup $\langle \rho_1(J_2)\rangle$ is dense in the real torus $\{e^{itX+isY}\mid t,s\in\BR\}$. Set $D=X+Y=\mathrm{diag}(2,-4,1)$ and observe that all differences $D_{ii}-D_{jj}$ for $i\neq j$ are distinct and nonzero. 
		
		Given that $iX\in \mathfrak{k}$, we also have that $\mathrm{Ad}_{\rho_1(\si_1)}(iX)=i\rho_1(\si_1) X\rho_1(\si_1)^{-1}\in \mathfrak{k}$. Set $C=\rho_1(\si_1) X\rho_1(\si_1)^{-1}\in \mathfrak{k}_{\BC}$. We claim that $(D,C)$ acts irreducibly on $V=\hilb_1$. Observe that the basis vectors $i_x$ are eigenvectors for $D$ and the vectors $\rho_1(\si_1)i_x$ are eigenvectors for $C$. Any invariant subspace for the joint action is a sum of $D$-eigenspaces. Since all entries of $\rho_1(\si_1)$ are nonzero, $V$ is the only nonzero invariant subspace. 
		
        Corollary \ref{cor.su} implies that $\mathrm{SU}(M_1)\subseteq K$. To prove density on the whole unitary group fix $A\in\mathrm{U}(M_1)$. Notice that $\mathrm{tr}(Y)=-1$ so that $\mathrm{det}(e^{isY})=e^{-is}$. Fix $s$ so that $\mathrm{det}(e^{isY})=\mathrm{det}(A)$. Then $\widetilde{A}=Ae^{-isY}\in \mathrm{SU}(M_1)$ and $A=\widetilde{A}e^{isY}\in K$. Thus $K=\mathrm{U}(M_1)$.
	\end{proof}
	
	\subsection{Multiple qubit universality}
	
	We prove density of the braid group action on $\hilb_2^\mathrm{comp}$ and more generally $\hilb_N^\mathrm{comp}$. We do not prove a density theorem on the entire encoding space. 
	
	Let us first consider the braids $J_3=\si_3\si_4^2\si_3$ and $\si_2$ in $B_5$ which, in the representation $\rho_2$, fix the first and fourth letters on path $x$ for $i_x\in \hilb_2$. Let 
    \begin{align}
        \overline{\hilb}_2=\{gv\in \hilb_2\mid g \in \langle \rho_2(J_3),\rho_2(\si_2)\rangle, v\in \hilb_2^\mathrm{comp}\}.
    \end{align}
    The invariant subspaces of $\overline{\hilb}_2$ are as follows
	\be\label{eq.Vdecomp}
	\begin{aligned}
		V_{s\cdot\cdot u}&=\mathrm{span}\{i_x\mid x\in\{ssuu, susu, srlu\}\},
		&\qquad
		L_{suus}&=\mathrm{span}\{i_{suus}\},\\
		V_{u\cdot\cdot s}&=\mathrm{span}\{i_x\mid x\in\{usus, uuss, urls\}\},
		&\qquad
		L_{ussu}&=\mathrm{span}\{i_{ussu}\}.
	\end{aligned}
	\ee
	The actions may be described in terms of the group action on $\hilb_1$.
	In particular, $\rho_2(J_3)$ and $\rho_2(\si_2)$ act by scalars on $L_{suus}$ and $L_{ussu}$.
	
	\begin{lemma}\label{lem.shift}
		The action of $\langle J_3,\si_2\rangle$ on $\overline{\hilb}_2$ preserves the components of \eqref{eq.Vdecomp}. Moreover,
		\begin{align*}
			(\rho_2(J_3),\rho_2(\si_2))\vert_{s\cdot\cdot u}&=(\rho_1(J_2),\rho_1(\si_1))\vert_{\a\to\a-1/2},
			\\
			(\rho_2(J_3),\rho_2(\si_2))\vert_{u\cdot\cdot s}&=(\rho_1(J_2),\rho_1(\si_1))\vert_{\a\to\a+1/2}.
		\end{align*}
	\end{lemma}
	\begin{proof}
		The claim for $\rho_2(J_3)$ follows from naturality of the inclusion maps with the braiding. For $\rho_2(\si_2)$, we use the locality of the action.

        The parameter shifts are based on the the last path letter $w_4$ which defines an inclusion $i_{w_4}:\bV\to \bW\otimes V(n_{3},\a_{3})$ and determines the parameter shift:
        \[\a_{3}=\begin{cases}
            \a+1/2 & w_4=s\\
             \a-1/2 & w_4=u
        \end{cases}.\qedhere\]
	\end{proof}
	
	Write $M_{s\cdot\cdot u}$ and $M_{u\cdot\cdot s}$ for the restriction of $M_2$ to the corresponding subspaces $ V_{s\cdot\cdot u}$ and $ V_{u\cdot\cdot s}$. These are corresponding shifts of the Hermitian pairing $M_1$. Also write
    \begin{align}
        M_N^{\mathrm{comp}}=M_N\vert_{\hilb_N^{\mathrm{comp}}}\,.
    \end{align}

    We introduce a further restricted parameter space describing pairs $(\th, \a)$ for which positivity of quantum numbers holds on the shifts $\a+\frac{1}{2}$ and $\a-\frac{1}{2}$:
    \begin{align}\label{eq.R'}
        \mathcal{R}'=\left\{ (\th,\a)\in \BR^2\mid 0<\th<\frac{\pi}{2},\quad 1<\a<\frac{\pi}{\th}-2\right\}.
    \end{align}
	
	\begin{theorem}\label{thm.twoqubit}
		Suppose $(\th, \a)\in \mathcal{R}'$ satisfy the $\BZ$-independence condition \eqref{eq.generic}. Then the closure of the action of $\langle \rho_2(J_3), \rho_2(\si_2)\rangle$ on $\overline{\hilb}_2$ contains \[
		\mathrm{SU}(M_{s\cdot\cdot u})\times \mathrm{SU}(M_{u\cdot\cdot s})\times \{1\}\times \{1\}.
		\]
	\end{theorem}
	\begin{proof}
		We approach the proof similar to that of Theorem \ref{thm.onequbit}. Set $W=V_{s\cdot\cdot u}\oplus V_{u\cdot\cdot s}$ and $L=L_{suus}\oplus L_{ussu}$. By Lemma \ref{lem.shift}, $\rho_2(J_3)\vert_{s\cdot\cdot u}=e^{i\th(\a X_{s\cdot\cdot u}+Y_{s\cdot\cdot u})}$ and $\rho_2(J_3)\vert_{u\cdot\cdot s}=e^{i\th(\a X_{u\cdot\cdot s}+Y_{u\cdot\cdot s})}$ where
        \begin{align*}
            X_{s\cdot\cdot u}&=X_1=\mathrm{diag}(2,-2,0), & Y_{s\cdot\cdot u}&=(Y_1-\frac{1}{2}X)=\mathrm{diag}(-1,-1,1),
            \\
            X_{u\cdot\cdot s}&=X_1=\mathrm{diag}(2,-2,0), & Y_{u\cdot\cdot s}&=(Y_1+\frac{1}{2}X)=\mathrm{diag}(1,-3,1).
        \end{align*}
		Then $\rho_2(J_3)\vert_W=\rho_2(J_3)\vert_{s\cdot\cdot u}\oplus \rho_2(J_3)\vert_{u\cdot\cdot s}$ and we write $X_2=X_{s\cdot\cdot u}\oplus X_{u\cdot\cdot s}$ and $Y_2=Y_{s\cdot\cdot u}\oplus Y_{u\cdot\cdot s}$. Again, $\BZ$-linear independence implies that $\langle \rho_2(J_3)\vert_{W} \rangle$ is dense in $\{e^{itX_2+isY_2}\mid t,s\in\BR\}$. Set $D=X_2+4Y_2$ so that all differences of weights are nonzero and distinct. Set $C=\rho_2(\si_2)\vert_{W} X_2\rho_2(\si_2)\vert_{W}^{-1}$, which differs from two copies of $C$ defined in the proof of Theorem \ref{thm.onequbit} by parameter shifts. 
        
        Extend the actions of $D$ and $C$ to $W\oplus L$, writing $\widehat D$ and $\widehat C$ for these actions. Since $L$ is a sum of 1-dimensional spaces with respect to the actions $\rho_2(J_3)$ and $\rho_2(\si_2)$, and thus $\widehat D$ and $\widehat C$, their actions commute on $L$. The theorem follows from Corollary \ref{cor.su} applied to the action of $(\widehat D, \widehat C)$ on $\overline{\hilb}_2$.
	\end{proof}

    \begin{corollary}
        Suppose that $(\th,\a)$ satisfy the hypothesis of Theorem \ref{thm.twoqubit}. Then braiding approximates any operator in $\mathrm{U}(M_2^{\mathrm{comp}})$.
    \end{corollary}
    \begin{proof}
        By Theorem \ref{thm.twoqubit}, we can approximate the unitary operator
		\[
		1_{V_{s\cdot\cdot u}}\times \mathrm{diag}(e^{i\varphi},e^{-i\varphi},1)\times 1_{L_{suus}}\times 1_{L_{ussu}}
		\]
		on $\overline{\hilb}_2$. Now, restricting to the computational subspace $\{i_x\mid x\in\{susu,suus,ussu,usus\}\}$ yields the operator $\mathrm{diag}(1,1,1,e^{i\varphi})$. In particular, it converges with arbitrarily small leakage. Moreover, naturality along with Theorem \ref{thm.onequbit} allows us to construct arbitrary single-qubit gates on either qubit.

        By the Brylinski--Brylinski theorem \cite[Thm.1.3]{BrylinskiBrylinski}, the set of single qubit gates along with any imprimitive (entangling) unitary, such as $CZ=\mathrm{diag}(1,1,1,e^{i\varphi})$ at $\varphi=\pi$,  form a universal set of gates.
    \end{proof}
	
	The universality result for $n$ qubits now follows from naturality of the braiding with respect to the inclusion maps. More precisely, the one- and two-qubit approximations apply to operators on any single qubit and consecutive pair of qubits in $\hilb_N$. It is also worth noting that $M_N^{\mathrm{comp}}$ is positive definite on every intermediate computational channel as the only possible parameter values are $\a$ and $\a\pm \frac{1}{2}$ with $(\th,\a)\in\mathcal{R}'$. Again by  \cite[Thm.1.3]{BrylinskiBrylinski}, we have the main result.
	
	\begin{corollary}\label{cor.univ}
		Suppose that $\th$ and $\a$ satisfy the hypotheses of Theorems \ref{thm.onequbit} and \ref{thm.twoqubit}. For each $N\geq 1$, braiding implements universal computation on $\hilb_N^\mathrm{comp}$.
	\end{corollary}

    \section*{Tool Use}No AI tools have been used to write or create this work.
	\bibliographystyle{hamsalpha}
	\bibliography{biblio}
\end{document}